\documentclass[12pt,twoside]{article}   
\usepackage[super,sort,comma]{natbib}

\usepackage{fancyhdr}		

\usepackage[section]{placeins}   %

\usepackage{url}
\usepackage{caption} 
\usepackage{framed,multirow}

\usepackage{amssymb}
\usepackage{amsmath}
\usepackage{latexsym}

\usepackage{color, soul}
\usepackage{xargs}
\usepackage{tikz}
\usepackage{todonotes}
\usepackage{multirow} 
\usepackage{verbatim}
\usepackage{nameref,hyperref}

\usepackage{url}
\usepackage{xcolor}

\usepackage{hyperref}

\usepackage{graphicx}

\makeatletter \renewcommand\@biblabel[1]{$^{#1}$} \makeatother
\renewcommand{\thesection}{{\sf \Roman{section}}.}
\renewcommand{\thesubsection}{\thesection{\sf \Alph{subsection}}.}
\renewcommand{\thesubsubsection}{\thesubsection{\sf \arabic{subsubsection}}.}
\renewcommand{\theparagraph}{\alph{paragraph}.}

\newcommand{\cen}[1]{\begin{center} #1 \end{center}}

\usepackage[mathlines]{lineno}

\usepackage{hyperref}
\hypersetup{ colorlinks,
    citecolor=blue,
    filecolor=blue,
    linkcolor=blue,
    urlcolor=blue
}

\usepackage{xcolor}

\definecolor{gray}{rgb}{0.6,0.6,0.6}
\definecolor{red}{rgb}{0.85,0,0}
\definecolor{green}{rgb}{0,0.85,0}
\definecolor{blue}{rgb}{0,0,0.85}
\definecolor{beige}{rgb}{0.92,0.87,0.78}
\usepackage[all]{hypcap}    

\begin{document}


\cen{\sf {\Large {\bfseries {Progressively refined deep joint registration segmentation (ProRSeg) of gastrointestinal organs at risk: Application to MRI and cone-beam CT}}}  \\  
\vspace*{10mm}
\small{Jue Jiang$^{1,\dagger}$, Jun Hong$^{1,\star}$, Kathryn Tringale$^2$, Marsha Reyngold$^2$, Christopher Crane$^2$, Neelam Tyagi$^1$, Harini Veeraraghavan$^{1,\dagger}$}} Department of Medical Physics$^{1}$, Memorial Sloan Kettering Cancer Center, 1275 York Avenue, New York, NY 1006 \\
Department of Radiation Oncology$^{2}$, Memorial Sloan Kettering Cancer Center, 1275 York Avenue, New York, NY 1006 \\
$\dagger$: These authors contributed equally \\
$\star$: work performed when author was at MSKCC \\
Corresponding Author Address: Box 84 - Medical Physics, Memorial Sloan Kettering Cancer Center, 1275 York Avenue, New York, NY 10065   .\\
Corresponding Author Email: veerarah@mskcc.org \\

\begin{abstract}

{\bf{Background}: \/}\rm Adaptive radiation treatment (ART) for locally advanced pancreatic cancer (LAPC) requires consistently accurate segmentation of the extremely mobile gastrointestinal (GI) organs at risk (OAR) including the stomach, duodenum, large and small bowel. Also, due to lack of sufficiently accurate and fast deformable image registration (DIR), accumulated dose to the GI OARs is currently only estimated, further limiting the ability to more precisely adapt treatments. \\
{\bf Purpose: \/}\rm Develop a 3-D  \underline{pro}gressively refined joint \underline{r}egistration-\underline{seg}mentation (ProRSeg) deep network to segment and align treatment MRIs, then evaluate segmentation accuracy, registration consistency, and feasibility for OAR dose accumulation.   \\
{\bf Method: \/}\rm ProRSeg was trained using 5-fold cross-validation with 110 T2-weighted MRI acquired at 5 treatment fractions from 10 different patients, taking care that same patient scans were not placed in training and testing folds. Segmentation accuracy was measured using Dice similarity coefficient (DSC) and  Hausdorff distance at 95th percentile (HD95). Registration consistency was measured using coefficient of variation (CV) in displacement of OARs. Ablation tests and accuracy comparisons against multiple methods were done. Finally, applicability of ProRSeg to segment cone-beam CT (CBCT) scans was evaluated on 80 scans using 5-fold cross-validation.    \\
\textbf{Results: \/}\rm ProRSeg processed 3D volumes (128 $\times$ 192 $\times$ 128) in 3 secs on a NVIDIA Tesla V100 GPU. It's segmentations were significantly more accurate ($p<0.001$) than compared methods, achieving a DSC of 0.94 $\pm$0.02 for liver, 0.88$\pm$0.04 for large bowel, 0.78$\pm$0.03 for small bowel and 0.82$\pm$0.04 for stomach-duodenum from MRI. ProRSeg achieved a DSC of 0.72$\pm$0.01 for small bowel and 0.76$\pm$0.03 for stomach-duodenum from CBCT. ProRSeg registrations resulted in the lowest CV in displacement (stomach-duodenum $CV_{x}$: 0.75\%, $CV_{y}$: 0.73\%, and $CV_{z}$: 0.81\%; small bowel $CV_{x}$: 0.80\%, $CV_{y}$: 0.80\%, and $CV_{z}$: 0.68\%; large bowel $CV_{x}$: 0.71\%, $CV_{y}$ : 0.81\%, and $CV_{z}$: 0.75\%). ProRSeg based dose accumulation accounting for intra-fraction (pre-treatment to post-treatment MRI scan) and inter-fraction motion showed that the organ dose constraints were violated in 4 patients for stomach-duodenum and for 3 patients for small bowel. Study limitations include lack of independent testing and ground truth phantom datasets to measure dose accumulation accuracy.  \\
{\bf Conclusions: \/}\rm ProRSeg produced more accurate and consistent GI OARs segmentation and DIR of MRI and CBCTs compared to multiple methods. Preliminary results indicates feasibility for OAR dose accumulation using ProRSeg. 
 \end{abstract}
\normalsize

\section{Introduction}
MR-guided adaptive radiation therapy (MRgART) is a new treatment that allows for radiative dose escalation of locally advanced pancreatic cancers (LAPC) with higher precision than conventionally used cone-beam CTs (CBCT) due to improved soft-tissue visualization on MRI. MR-LINAC treatments also allow daily treatment adaptation and replanning to account for the changing anatomy. Anatomy changes result from day-to-day variations in organ shape and configuration as well as motion due to peristalsis and breathing, all of which introduce large geometric uncertainties to the delivery of radiation. However,  widespread adoption of MRgART is hampered by the need for manual contouring and plan re-optimization, which together can take 40 to 70 mins\cite{Hurst2021,henke2018} daily. Hence, there is a clinical need for consistently accurate and fast auto-segmentation of the gastrointestinal (GI) organs at risk (OARs) including the stomach, duodenum, small and large bowel.
\\
Highly accurate segmentation, measured as a Dice similarity coefficient (DSC) exceeding 0.8 of abdominal organs such as the liver, kidneys, spleen, as well as the stomach (excluding duodenum) has been reported by using off-the-shelf deep learning (DL) architectures including nnUnet\cite{isensee2021nnu} and Unet\cite{ronneberger2015u}, as well as customized dense V-Nets\cite{Gibson2018TMI} and new transformer based methods\cite{chen2020} applied to CT images. Slice-wise priors provided as manual segmentations\cite{zhang2022}, multi-view methods using inter-slice information from several slices and dense connections\cite{chen2020} as well as self-supervised learning of transformers \cite{jiang2022self} have shown the ability to segment the more challenging GI OARs such as large and small bowel and duodenum from MRI. However, the need for manual editing\cite{zhang2022} and large number of adjacent slices\cite{chen2020} required to provide priors may reduce the number of available training sets and the practicality (due to need for manual editing) of such methods. 
\\
Besides segmentation, reliable deformable image registration (DIR) is also needed for voxel-wise OAR dose accumulation in order to ensure that the prescription dose was delivered to the targets while sparing organs of unnecessary radiation. DIR based contour propagation\cite{velec2017,van2019mri,zhang2020} is one approach to simultaneously solve both deformable dose accumulation and segmentation. This is also a convenient option as DIR methods are commonly available techniques in commercial software platforms. However, commonly available DIR methods often use small deformation frameworks based on parameterizing a displacement field added to an identity transform, which cannot preserve topology\cite{Ashburner2007} for large organ displacements. Deep learning image registration (DLIR) methods\cite{xu2019deepatlas,he2020deep,estienne2019u,balakrishnan2019voxelmorph} are often faster and more accurate than iterative registration methods because they directly compute the diffeomorphic transformation between images in a single step instead of solving a non-linear optimization to align every image pair. DLIR methods, which typically use stationary velocity field (SVF) to compute the diffeomorphic transformation, reduce the computational requirements by reducing the search space to a set of diffeomorphisms that are within a Lie structure, but it also limits their flexibility in handling large and complex deformations\cite{mok2020fast}. Furthermore, DLIR network optimization as well as iterative registration optimization typically focuses on minimizing an energy function composed of global smoothness and a variational intensity regularization, which is insufficient to handle abrupt and large differences in motion occurring between organs, especially at the organ boundaries\cite{Fu2018}. 
\\
Compositional DIR strategies that extract diffeomorphic transformation in stages such as used for non-sliding and sliding organs\cite{Fu2018}, adaptive anisotropic filtering of the incrementally refined deformation vector field (DVF)\cite{papiez2014} as well as cascaded network formulations\cite{deVos2019MedIA,zhao2019recursive,Ashburner2007} are more robust than the single step methods for handling large deformations while still retaining the SVF assumption, thus providing computational speed up compared to the time-dependent diffeomorphic registration methods. However, cascaded DLIR methods are limited by the memory requirements and thus require sequential training of individual networks, which increases training time. Additionally, because the networks in the cascade are trained one after another, there is no guarantee that the deformations modeled in the prior steps will be retained in the future steps. Recurrent registration method (R2N2) that computes local parameterized Gaussian deformations\cite{sandkuhler2019recurrent} has demonstrated ability to model large anatomic deformations occurring in a respiratory cycle. However, the use of local parametrization restricts the flexibility of this approach to handle large and continuous deformations. R2N2 was shown to be less accurate than a progressive registration method computing a continuous deformation flow field for quantifying longitudinal tumor volume changes\cite{jiang2022CBCTTMI}. Our approach improves on these works to compute topology preserving (quantified by non-negative Jacobian determinant) diffeomorphic deformations and multi-organ segmentations using a progressive joint registration-segmentation (ProRSeg) approach, wherein deformation flow computed at a given step is conditioned on the prior step using a 3D convolutional long short term memory network (CLSTM)\cite{shi2015convolutional}. Because the DVF is modeled as a continuous and differentiable flow-field, it is invertible, thus ensuring diffeomorphic transformations. ProRSeg is optimized using a multi-tasked learning of a registration and segmentation network, which allows it to leverage the implicit backpropagated errors from the two networks. Multi-tasked networks have previously shown to produce more accurate normal tissue segmentation than individually trained DL networks\cite{xu2019deepatlas, he2020deep, estienne2019u,beljaardsCrossStitch2020}.  
\\
ProRSeg is most similar to a prior registration-segmentation method that we developed for tracking lung tumors\cite{jiang2022CBCTTMI} from cone-beam CT (CBCT) images. However, ProRSeg accounts for both respiratory and large organ shape variations, while our prior work was only concerned with tracking linearly shrinking tumors during radiation treatment. ProRSeg aligns images with large deformation by computing a smooth interpolated sequence of dense deformation flows using 3D CLSTM\cite{shi2015convolutional} networks implemented in the encoder layers of registration and segmentation networks. The CLSTM explicitly enforces consistency between the individual steps and conditions the deformations computed in subsequent steps on the prior steps, which allows incremental refinement of spatial alignment without destroying prior alignment. CLSTM formulation employs convolutional filters, thereby allowing for spatially continuous and differentiable DVF formulation. Second, the segmentation network uses progressively aligned spatial appearance and geometry priors pertaining to previous treatment fraction (produced by the registration network) as inputs to constrain the segmentations. Hence, the segmentation network avails information about the organs and their geometry from a prior treatment fraction, which increases robustness to arbitrary variations occurring in the GI organs. In contrast, prior works\cite{xu2019deepatlas,estienne2019u} used a weaker regularizing constraint to ensure that the outputs of registration and segmentation networks matched as additional losses during training. Third, segmentation consistency loss enforcing similarity of segmentation produced by registration network, the segmentation network and the expert provides supervised feedback to both networks. We show such a loss improves accuracy.
\\
\begin{figure*}[t]
	\begin{center}
        \includegraphics[width=0.9\columnwidth,scale=0.1]{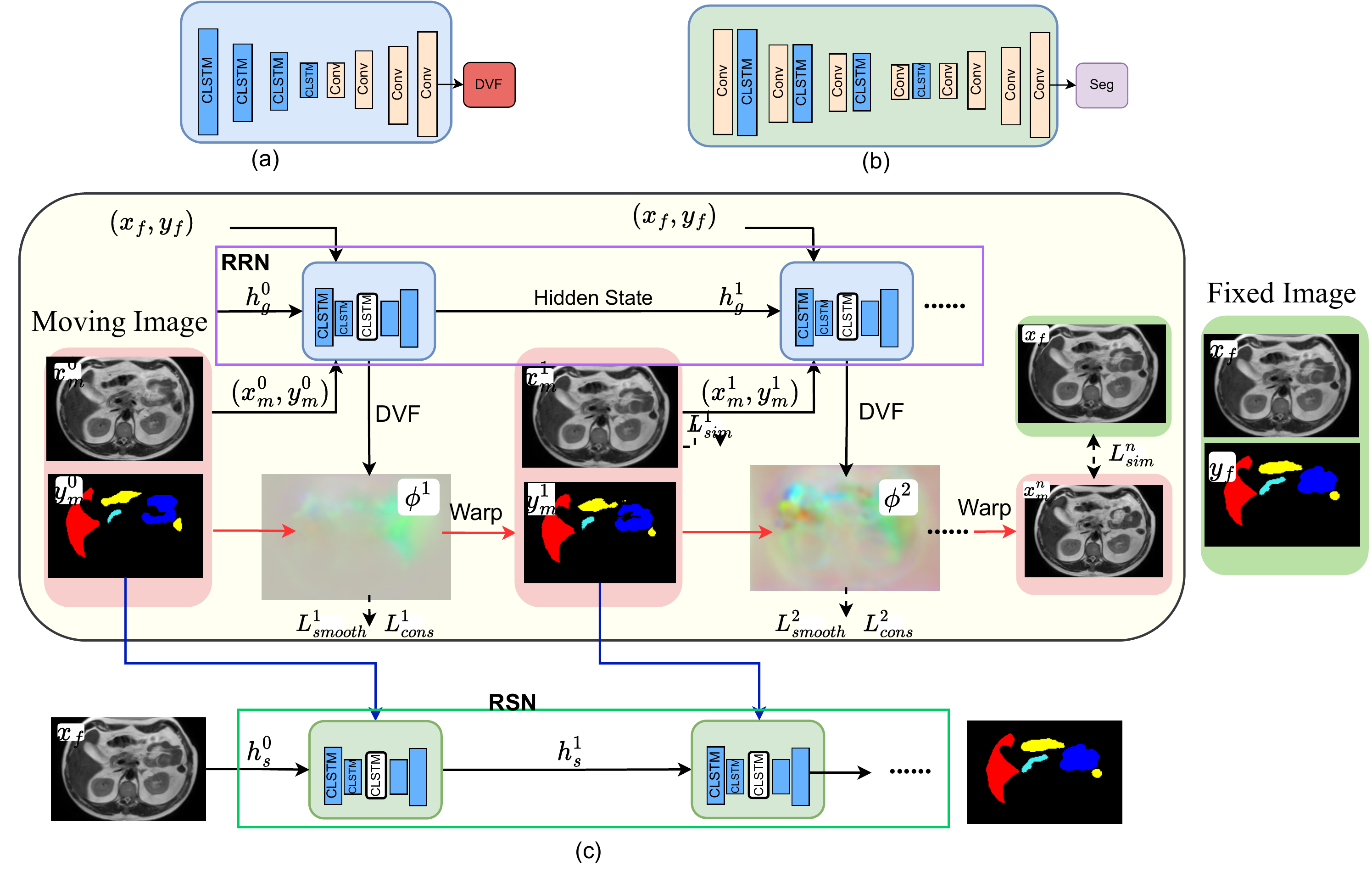}
		\caption{\small (a) Schematic of recurrent Registration Network (RRN), where convolutional (Conv) layers in the encoder are combined with 3D-CLSTM. (b) Recurrent Segmentation Network (RSN) uses a Unet-3D backbone with 3D-CLSTM placed after convolutional blocks in the encoder layers. (c) ProRSeg combines RRN and RSN. The unrolled representation showing CLSTM in the encoder layers for progressively refining the registration and segmentation are shown. RSN combines $x_t$ with the progressively aligned images $x_{m}^{i=1,\ldots N}$ and segmentations $y_{m}^{i=1,\ldots N}$ produced by RRN as inputs to its CLSTMs to generate segmentation $y_{t}$ in N steps.  
		\label{fig:overview_method}}
		\end{center}
\end{figure*}
\underline{Our contributions are\/}\rm: (i) a simultaneous registration-segmentation approach for segmenting GI OARs from MRI while computing voxel-wise deformable dose accumulation, (ii) use of registration derived spatially aligned appearance and geometry priors to constrain segmentation that increases accuracy, (iii) use of a 3D CLSTM implemented in the encoders of both registration and segmentation networks that increases robustness to arbitrary organ deformations by modeling such deformations as a progressively varying dense flow field. (iv) Finally, we evaluated ProRSeg for segmenting GI organs (stomach-duodenum and small bowel) from treatment CBCTs. 
\section{Materials and Method}
\subsection{Pancreas MRI dataset for MR-MR registration-segmentation}
The retrospective analysis was approved by the institutional internal review board. One hundred and ten 3D T2-weighted MRIs acquired from on treatment MRIs from 10 patients undergoing five fraction MR-guided SBRT to a total dose of 50 Gy were analyzed. A pneumatic compression belt set according to the patient convenience was used to minimize gross tumor volume (GTV) and GI organs motion occurring within 5mm of the GTV as described in our prior study\cite{tyagi2021}. The dose constraints to GI organs were defined as D$_{max}$ or D$_{0.035 cm3}$ $\leq$ 33 Gy and D$_{5cm3}$ $\leq$ 25 Gy. D$_{5cm3}$ for large bowel was $\leq$ 30 Gy. In each treatment fraction, three 3D T2-weighted MRI (TR/TE of 1300/87 ms, voxel size of 1 $\times$ 1 $\times$ 2 mm3, FOV of 400 $\times$ 450 $\times$ 250 mm3) were acquired at pre-treatment, verification, and following treatment on that fraction called post-treatment MRI. Verification MRI was acquired immediately before beam on to confirm patient anatomy had not shifted during contouring and adaptive replanning. Six patients had pre-treatment, verification, and post-treatment MRI with segmentation on all five fractions with the remaining 4 containing only pre-treatment MRI for all fractions. Additional details of treatment planning are in prior study\cite{tyagi2021,Alam2022}. 
\\
\textbf{Expert contouring details: \/}\rm Stomach-duodenum, large bowel, and small bowel, as well as liver were contoured on all the available treatment fraction MRIs by an expert medical student and verified by radiation oncologists, and represented the ground truth for verifying the ProSeg segmentations and deformable image registration (DIR).  

\subsection{Pancreas dataset for pCT-CBCT registration-segmentation}
ProRSeg was additionally evaluated on 80 CBCT scans acquired from 40 patients with LAPC and treated with hypofractionated RT on a regular linac (15 to 20 fractions with daily CBCTs used for guidance) from an institutional IRB approved retrospective research protocol and used in a prior work\cite{han2021}. For the purpose of study, a planning CT (pCT) and two CBCT scans acquired on different days in a deep inspiration breath hold (DIBH) state using an external respiratory monitor (Real-time Position Monitor, Varian Medical Systems) were used. pCT scans were acquired in DIBH with a diagnostic quality scanner (Brilliance Big Bore, Philips Health Systems; or DiscoveryST, GE Healthcare). The kilovoltage CBCT scans were acquired with 200-degree gantry rotation. The CBCT reconstruction diameter was 25 cm and length was 17.8 cm. 
\\
\textbf{Expert contouring details:\/} \rm Both pCT and CBCT scans were delineated by a radiation oncologist to provide stomach and first two segments of the duodenum and the remainder of the small bowel. On the CBCTs, only portion of the OARs within a volume of interest defined as 1cm expansion on 3D including the high-dose and ultra-high dose planning target volume (PTV)\cite{han2021}. 
\subsection{Image preprocessing details}
All scans were rigidly aligned with the prior treatment fraction MRI to bring images to the same spatial coordinates using methods available in the CERR software\cite{apte2012}. The CBCT scans were rigidly aligned to the planning CT (pCT) scans. MR images were also standardized using N4 bias field correction and histogram standardization to a randomly chosen reference using functions implemented into the CERR software. Finally, images were automatically cropped to the body region in order to remove table and air outside of the body using intensity thresholding, hole filling, followed by largest connected component extraction. 

\subsection{ProRSeg: Progressively refined joint registration segmentation}
\subsubsection{Approach Overview:}
ProRSeg is implemented using 3D convolutional recurrent registration or RRN ($g$) and recurrent segmentation networks or RSN ($s$). RRN uses a pair of source and target images $\{x_{m}, x_{f}\}$ and computes a dense deformation flow field to warp the source image into target ($x_{f}$) image's spatial coordinates (or $x_{m}^{f}$) by using a progressive sequence of deformations ($x_{m}^{f} = \{x_{m}^{1}, \ldots x_{m}^{N}\}$), where $N$ is the number of 3D CLSTM steps. The 3D CLSTM is implemented into all the encoder layers of RRN and RSN. The RSN generates a segmentation for $x_{f}$ by combining $x_f$ with progressively warped moving images and contours $y_{m}$ produced by the RRN ($\{x_{m}^{1},y_{m}^{1}\},\ldots \{x_{m}^{N},y_{m}^{N}\}$) as inputs to each CLSTM step (Figure~\ref{fig:overview_method}). 
\subsubsection{Convolutional long short term memory network (CLSTM):\/}\rm CLSTM is a type of recurrent neural network, which maintains long term contextual information about the state $x_{t}$ at step $t$ by using gating filters called forget gate $f^{t}$ and memory cells $c^{t}$, implemented using sigmoid activation function and a multiplicative term (Eqn \ref{eqn:CLSTM}). CLSTM improves upon long-short term memory network by using convolutional filters to maintain the state information using a dense encoding of the spatial neighborhood or the whole image. The CLSTM components including the state, forget gate, memory cells, hidden state $h^{t}$, input state $i^{t}$, and output gate $o^{t}$ are updated as below: 
\begin{equation}
\begin{split}
\label{eqn:CLSTM}
f^{t}&=\sigma(W_{xf}*x^{t}+W_{hf}*h^{t-1}+b_{f})\\
i^{t}&=\sigma(W_{xi}*x^{t}+W_{hi}*h^{t-1}+b_{i})\\
\tilde{c}^{t}&=tanh(W_{x\tilde{c}}*x^{t}+W_{h\tilde{c}}*h^{t-1}+b_{\tilde{c}})\\
o^{t} &= \sigma(W_{xo}*x^{t}+W_{ho}*h^{t-1}+b_{o})\\
c^{t} &= f^{t} \odot c^{t-1}+i^{t}\odot \tilde{c}^t\\
h^{t} &= o^{t} \odot tanh(c^{t}),
\end{split}
\end{equation}
where, $\sigma$ is the sigmoid activation function, $*$ the convolution operator, $\odot$ the Hadamard product, and $W$ the weight matrix.

\subsubsection{Recurrent registration network:}
A schematic of the RRN $g$ architecture is depicted in Figure~\ref{fig:overview_method} (a). RRN deforms an image $x_m$ into $x_{m}^{f}$, expressed as $g(x_{m}, x_{f}): \theta_{g}(x_m) \rightarrow x_{m}^{f}$ by computing a sequence of progressive deformation vector fields (DVF) using $N > 1$ CLSTM steps: $\phi_{m}^{f} = \phi^{1}  \circ \phi^{2}   ... \circ \phi^{N}$. $\phi^{i} : I + u^{i}$, where I is the identity and $u$ is the DVF. The input to the first layer is a channel-wise concatenated pair of source and target images ($\{x_{m}, x_{f}\}$) and the hidden state $h_{g}^{0}$ initialized to 0. Subsequent layers use the progressively deformed source $x_{m}^{i-1}$ and the hidden state $h_{g}^{i-1}$ output from the prior CLSTM step $i-1$ together with the target image $x_{f}$ as inputs to the current CLSTM step $i$. Images are channel-wise concatenated ($\{x_{m}^{i-1}, x_{f}\}$) for use in the CLSTM step. A CLSTM step $i$ computes a warped image and contour ($y_m^{i}$) as:
\begin{equation}
\setlength{\abovedisplayskip}{1pt}
\setlength{\belowdisplayskip}{1pt} 
\begin{split}
x_{m}^{i} = x_{m}^{i-1} \circ \phi^{i} \\
y_{m}^{i} = y_{m}^{i-1} \circ \phi^{i}.
\end{split}
\label{eqn:reg_phi1}
\end{equation}
Note that the contour $y_m$ is not used as an input to constrain the RRN network.
\\
RRN is optimized without any ground truth DVFs. Deep image similarity $L_{sim}$ and deep smoothness losses $L_{smooth}$ are used to regularize the warped image $y_{m}^{i}$ and the DVF $\phi^{i}$ of each CLSTM step. A supervised segmentation consistency loss $L_{cons}$ comparing the warped contour $y_{m}^{f}$ produced after $N$ CLSTM steps of the RRN with the expert delineation $y_{f}$ was computed by measuring contour overlaps using Dice similarity coefficient (DSC):
\begin{equation}
    \begin{split}
    \setlength{\abovedisplayskip}{1pt}
    \setlength{\belowdisplayskip}{1pt}
    L_{cons} =\underset{i=0}{\overset{N}\sum}L_{cons}^i=1-\underset{i=0}{\overset{N}\sum} DSC(y_{f},g(x^{i}_m, y^{i}_m,x_{m},h^{i}_{g})).
    \label{eqn:reg_cons}
    \end{split}
\end{equation}
$L_{sim}$ is computed by comparing the warped images in each CLSTM step with the fixed images by using mean square error (MSE) loss for the MR to MR registration. In the case of pCT-CBCT registration experiment, $L_{sim}$ was computed using normalized Cross-Correlation (NCC) computed locally using window of 5$\times$5$\times$5 centered on each voxel to improve robustness to CT and CBCT intensity differences\cite{balakrishnan2019voxelmorph}. This can be expressed as: 
\begin{equation}
L_{sim} = \left \{
\begin{matrix}
 \sum_{i=1}^{N} MSE(x_{m}^{i},x_{f}) & \text{if MR to MR} \\
 \sum_{i=1}^{N} NCC(x_{m}^{i},x_{f}) & \text{if CT to CBCT} 
\end{matrix}
\right.
\end{equation}
The NCC loss at each CLSTM step $i$ is an average of all the local NCC calculations, thereby ensuring robustness to local variations. 
\\
$L_{smooth}$ was used to regularize the incremental deformation flow from each CLSTM step by averaging the flow field gradient within each voxel as:
\begin{equation}
\setlength{\abovedisplayskip}{1pt}
\setlength{\belowdisplayskip}{1pt} 
\begin{split}
L_{smooth} = \textcolor{black}{\underset{t=1}{\overset{N}\sum}\ L_{smooth}^t}= \underset{t=1}{\overset{N}\sum}\ \underset{p \in \Omega}{\sum}\ ||\nabla \phi^t(p)||^{2}/N.   
\end{split}
\end{equation}
 The total registration loss is then computed as:
 \begin{equation}
\setlength{\abovedisplayskip}{1pt}
\setlength{\belowdisplayskip}{1pt} 
\begin{split}
L_{reg} = L_{sim} + \lambda_{smooth}L_{smooth} + \lambda_{cons}L_{cons}, 
\end{split}
\end{equation}
where $\lambda_{smooth}$ and $\lambda_{cons}$ are tradeoff parameters.
\\
\textbf{Implementation details:\/ }\rm RRN was constructed by modifying the Voxelmorph (a 3D-Unet backbone) \cite{balakrishnan2019voxelmorph} such that the convolutional filters in the encoder were replaced with 3D-CLSTM. \textcolor{black}{Because the CLSTM extracts features by keeping track of prior state information, it computes features that capture both the temporal context and the dense spatial context (from the convolutional filters used to implement CLSTM)}. The last layer of the RRN was composed of a spatial transformation function based on spatial transform networks\cite{jaderberg2015spatial} to convert the stationary displacement field into DVF.
Diffeomorphic deformation was ensured by using a diffeomorphic integration layer\cite{dalca2019unsupervised} following the 3-D flow field output of the registration network $g$. More details of specific RRN network layers are in Supplemental Table 1. Eight 3D CLSTM steps were used for RRN as done in a different work applied to lung tumor segmentation from CBCT\cite{jiang2022CBCTTMI}. 

\subsubsection{Recurrent segmentation network:} 
Schematic of the RSN network $s$ is shown in Figure \ref{fig:overview_method}(b). RSN progressively refines the multi-class segmentation of a given target image $x_{f}$  using $N+1$ CLSTM steps. RSN uses one additional CLSTM than the RRN because the first CLSTM step uses the undeformed moving image $x_m$ and it's segmentation $y_m$ with the target image $x_t$ as channel-wise concatenated input. The remaining CLSTM steps use channel-wise concatenated input $\{x_{m}^{i}, y_{m}^{i},x_{f}\}$ where $x_{m}^{i}$ and $y_{m}^{i}$ are produced by the RRN CLSTM step $i$. Segmentation from each one of the RSN CLSTM steps are computed as $y_{f}^{i} = s(x^{i}_{m}, y^{i}_{m}, x_{f}, h_{s}^{i})$, where $h_{s}^{i}$ is the hidden state of the RSN CLSTM step $i$, 1 $\leq$ $i$ $\leq$ $N$ (Eqn.~\ref{eqn:reg_phi1}). 
\\
The RSN is optimized by computing a deep supervision segmentation loss comparing the segmentations produced after each CLSTM step with expert segmentation. This loss is computed using cross-entropy as:
\begin{equation}
    \begin{split}
    \setlength{\abovedisplayskip}{1pt}
    \setlength{\belowdisplayskip}{1pt}
    L_{seg} =\underset{t=0}{\overset{N}\sum}L_{seg}^t=\underset{t=0}{\overset{N}\sum} logP(y_{f}|s(x^{i}_t, y^{i}_t,x_{m},h^{i}_{m})).
    \label{eqn:segmentation_loss2}
    \end{split}
\end{equation}
The losses, $L_{seg}^{0}$,...$L_{seg}^{N-1}$ provide deep supervision to train RSN. 
\\
\textbf{Implementation details: \/}\rm RSN is implemented with a 3D Unet backbone with N+1 CLSTM steps implemented into the encoder layers. The standard 3DUnet was improved by replacing the first convolutional layer with a CLSTM before the max pooling layer. Each convolutional block was composed of two convolution units, ReLU activation, and max-pooling layer. This resulted in feature sizes of 32,64,128,256, and 512. Nine CLSTM steps were used to implement the RSN and GPU memory limitation was addressed using truncated backpropagation as used for RRN. The detailed network architecture for RRN and RSN are in Supplementary Table 1 and Supplementary Table 2. 
\subsection{Training details}
Both RRN and RSN are trained end-to-end and optimized jointly to use the losses computed from both networks for optimizing the networks parameters. The networks were implemented using Pytorch library and trained on Nvidia GTX V100 with 16 GB memory. The networks were optimized using ADAM algorithm with an initial learning rate of 2e-4 for the first 30 epochs and then decayed to 0 in the next 30 epochs and a batch size of 1. The $\lambda_{smooth}$ was set to 30  experimentally. 
\\
ProRSeg was trained separately for MR-to-MR and pCT-CBCT registration using five-fold cross-validation taking care that the same patient scans were not used in the training and corresponding validation folds. In order to increase the number of training examples, all possible pairs of images for each patient arising from different treatment fractions were used. In addition, online data augmentation using image rotation and translation was implemented to increase data diversity for training.

\section{Experiments and Results}
\subsection{Comparative experiments}
Representative segmentation only (Unet3D), registration-based segmentation using conventional iterative registration using large diffeomorphic metric mapping (SyN\cite{avants2008symmetric}), deep learning registration based segmentation using Voxelmorph\cite{balakrishnan2019voxelmorph}, as well as simultaneous registration-segmentation using UResNet\cite{estienne2019u} were trained using the same training and testing sets as ProRSeg with 3D image volumes of size 128 $\times$ 192 $\times$ 128. Image pairs were pre-aligned using rigid registration in order to bring them into similar spatial coordinates prior to application of the DIR methods. The CBCT OAR segmentation results of a published study using the same dataset\cite{han2021} are also shown to provide a reference comparison to our method.

\subsection{Ablation experiments}
Ablation experiments were done using the MRI dataset. Experiments were performed to study differences in accuracy when using RRN based segmentation versus when using RSN that combines information from RRN to compute the segmentation. The impact of spatially aligned appearance and shape prior provided by RRN to RSN and segmentation consistency loss on the segmentation accuracy were also measured.  Finally, the impact of number of CLSTM steps (1 to 8) on segmentation accuracy produced by RRN and RSN were analyzed.

\subsection{Metrics and statistical analysis:}
Segmentation accuracy was measured using the Dice similarity coefficient 
(DSC) and Hausdorff distance at 95th percentile (HD95) on the validation set (validation data not used for training in each cross-validation fold). Statistical comparison of accuracy differences between the various methods with respect to ProRSeg was measured using paired, two-sided Wilcoxon-signed rank tests at 95\% significance level. Only p values $< 0.05$ were considered significant. \\
Segmentation consistency was computed by using coefficient of variation (CV$_{DSC}$ \% = $\frac{\sigma_{DSC}}{\mu_{DSC}} \times 100$), where $\sigma_{DSC}$ is the standard deviation of the DSC per patient and $\mu_{DSC}$ is the population mean DSC. Furthermore, the variability in segmentation accuracy across treatment fractions was analyzed by measuring statistical differences in DSC and HD95 for the GI OARs extracted at 5 different treatment fractions for MRI using paired and two-sided Kruskal-Wallis tests at 95\% significance levels.  \\
Registration smoothness was measured using standard deviation of Jacobian determinant ($J_{SD}$) and the folding fraction $|J_{\phi}|$. Finally, consistency of registration to variations in anatomy was measured using percentage coefficient of variation (CV) in the median displacement for each GI OAR at patient level, by varying the source images (different treatment fractions) aligned to each treatment fraction MRI (as target). CV for each patient was evaluated in all three displacement directions as $CV = \frac{\sigma}{\mu}$, where $\sigma$ is the standard deviation and $\mu$ is the mean displacement.   

\subsection{GI OAR segmentation accuracy from MRI}
\begin{figure*}[h]
	\begin{center}
        \includegraphics[width=0.8\columnwidth,scale=0.1]{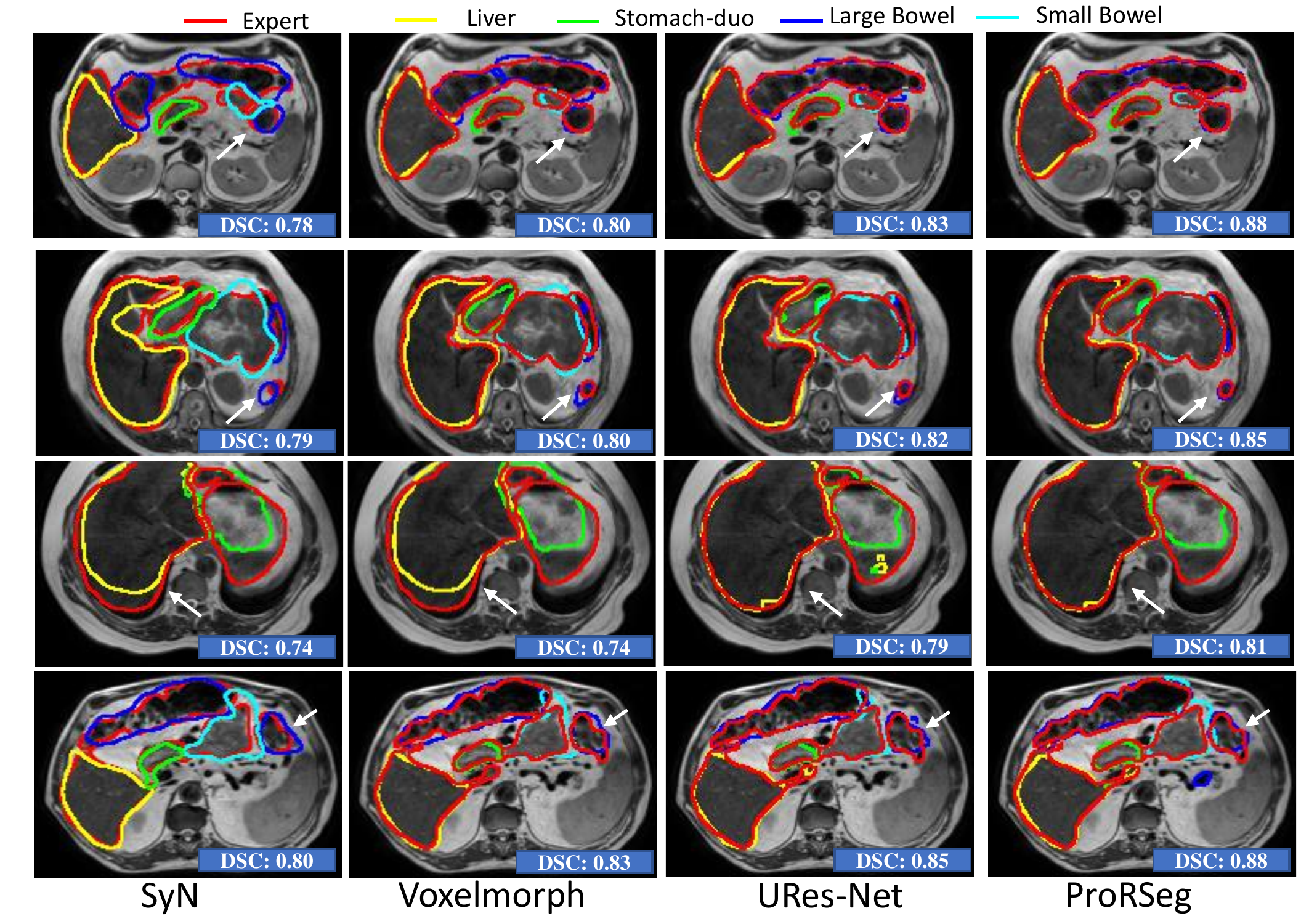}
		\caption{\small Comparison of OAR segmentations generated by multiple methods from MRI on representative examples.
			 \label{fig:seg_overlay_MRI}}
		\end{center}
\end{figure*}

\begin{figure*}
	\begin{center}
        \includegraphics[width=0.8\columnwidth,scale=0.1]{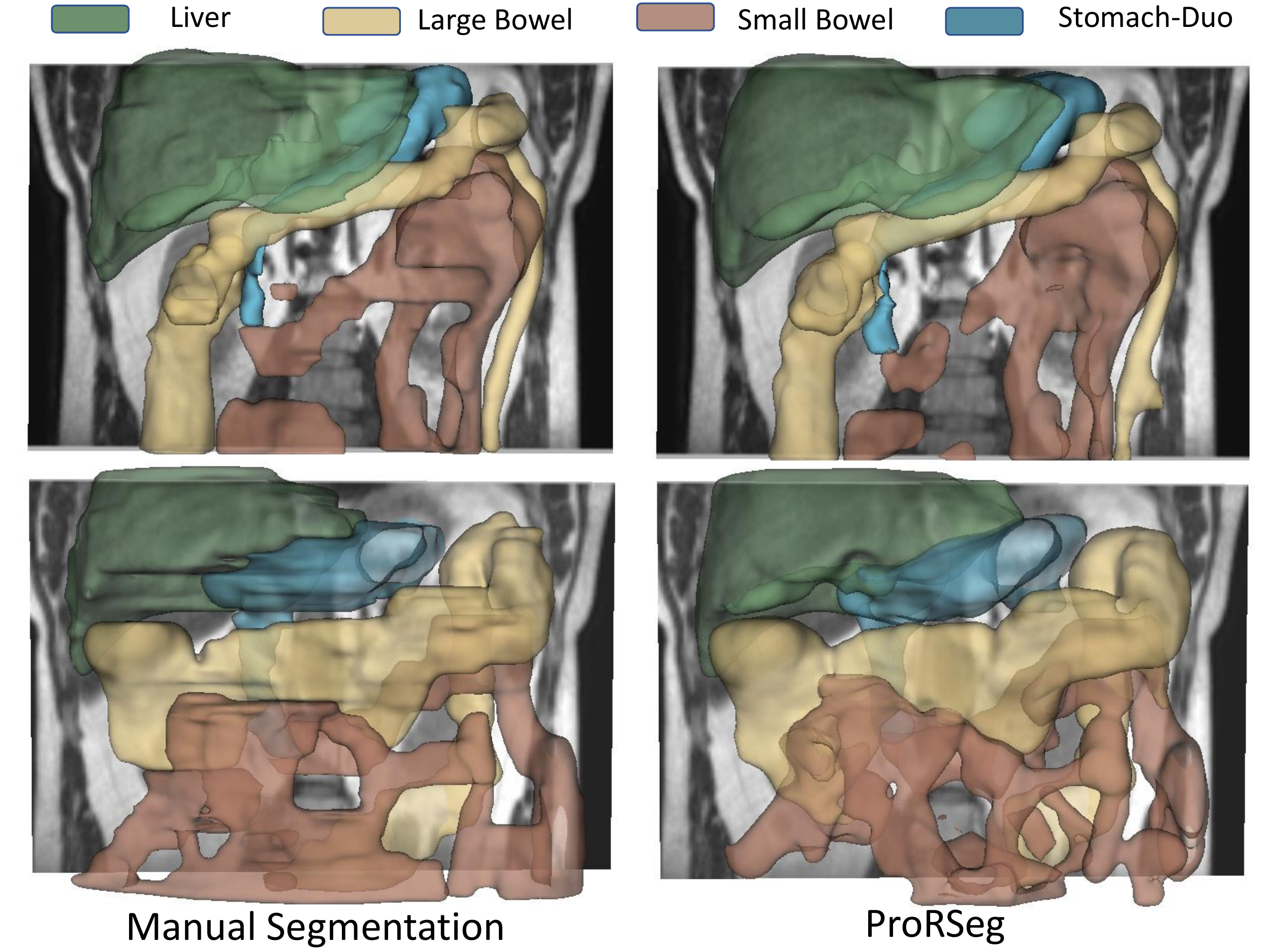}
		\caption{\small Three dimensional rendering of volumetric segmentations produced by expert (first column) and ProRSeg (second column) for best case (top row) and worst case (bottom row) patients.  
		\label{fig:3D_show}}
		\end{center}
\end{figure*}

Table \ref{tab:MRI_seg} shows the segmentation accuracies produced by the evaluated methods when aligning consecutive treatment fractions. ProRSeg produced the highest DSC and lowest HD95. ProRSeg was significantly ($p<0.001$) more accurate than the second best method, UNet3D (0.78 vs. 0.68) as well as the joint registration-segmentation UResNet\cite{estienne2019u} for all analyzed OARs ($p<0.001$). Supplemental Table 3 shows the p values measuring the differences between various methods with respect to ProRSeg. \\ Figure.~\ref{fig:seg_overlay_MRI} shows segmentation contours produced by the analyzed methods together with the expert delineations (in red) on representative examples. The overall DSC accuracy is also shown for all the cases and methods. As seen, ProRseg most closely matched the expert delineations even for hard to segment small bowel (Figure.~\ref{fig:seg_overlay_MRI} row 1, 2 and 4) and stomach-duodenum (Figure.~\ref{fig:seg_overlay_MRI} row 1, 2, 4). On the other hand, iterative registration\cite{avants2008symmetric} resulted in poor segmentations even for large organs such as the liver. Similarly, Voxelmorph\cite{balakrishnan2019voxelmorph}, a deep learning single step registration based segmentation method was unable to match the expert contours as closely as either the UResNet\cite{estienne2019u} or ProRSeg. Both UResNet and ProRSeg showed higher accuracy for the presented cases, except when large differences in organ shape and appearance occurred between treatment fractions. An example case with poor segmentation of the stomach occurring as a result of filled stomach aligned to empty stomach occurring in the prior treatment fraction is shown in Figure.~\ref{fig:seg_overlay_MRI}, row 3.
\\
Motivated by a prior study for MRI-based upper GI organs segmentation\cite{zhang2020} that fused DIR based segmentations from multiple prior fraction MRIs, ensemble segmentations were computed for the treatment fractions 4 and 5 by performing decision level fusion using simple majority voting of the segmentations produced by using multiple prior fraction MRIs (first to third fraction for treatment fraction 4, and first to fourth fraction for treatment fraction 5) as prior images for the registration-segmentation. This approach increased the segmentation accuracy for large bowel (0.91 $\pm$0.02 vs. 0.90 $\pm$ 0.04), small bowel (0.83$\pm$0.02 vs. 0.80 $\pm$0.05), and the stomach-duodenum (0.85$\pm$0.04 vs. 0.80$\pm$0.07). Ensemble segmentations were not generated for the first three fractions because at least three preceding treatment fraction MRIs were needed for meaningful segmentation fusion using majority voting.
\\
Figure~\ref{fig:3D_show} shows 3D rendering of two examples, the best case with an overall DSC of 0.88 and the worst case with an overall DSC of 0.81. As shown, for the best case example, ProRSeg closely matched the expert delineation of intra-peritoneal small bowel, and achieving a high DSC of 0.81 for small bowel. Reduction in overall accuracy in the second case occurred due to lower accuracy in segmenting the intra-peritoneal small bowel loops (DSC of 0.72). In comparison, the DSC for other OARs were high, stomach-duodenum DSC of 0.85 and large bowel DSC of 0.93. 
\begin{table} [t]
\centering{\caption{Segmentation accuracy (mean and standard deviation) of various methods applied to T2-w MRI. LG bowel: Large bowel, SM bowel: small bowel, Sto-Duo: stomach-duodenum. $^{\dagger}$: Segmentations generated for treatment fraction 4 and 5 using majority voting of segmentations produced by using  preceding treatment fractions as prior fractions.\\
$^{*}$: $p < 0.05$, $^{+}$: $p < 0.01$, $^{\ddagger}$: $p < 0.001$.}
\label{tab:MRI_seg} 
	\scriptsize
	\begin{tabular}{|c|c|c|c|c|c|c|c|c|} 
		\hline 
        \multirow{2}{*}{Method} & \multicolumn{4}{c|}{DSC $\uparrow$} & 
        \multicolumn{4}{c|}{HD95 mm $\downarrow$} \\
        \cline{2-9} 
		{}&{Liver}&{ LG Bowel}&{SM Bowel}&{Sto-Duo}&{Liver}&{ LG Bowel}&{SM Bowel}&{Sto-Duo} \\
		\hline
			{SyN\cite{avants2008symmetric}}&{0.89$\pm$0.04}&{0.59$\pm$0.13} & {0.61$\pm$0.09}&{0.66$\pm$0.08} &{9.17$\pm$3.55}&{20.04$\pm$8.55} & {20.74$\pm$7.36$^{\ddagger}$}&{13.08$\pm$5.08$^{\ddagger}$}\\
		\hline
		{Voxelmorph\cite{balakrishnan2019voxelmorph}}&{0.91$\pm$0.06}&{0.74$\pm$0.18} & {0.67$\pm$0.10}&{0.75$\pm$0.09} &{7.85$\pm$3.85}&{14.52$\pm$9.81} & {19.26$\pm$7.97$^{\ddagger}$}&{13.35$\pm$13.21$^{\ddagger}$}\\
		\hline
		{Unet3D}&{0.92$\pm$0.02 }&{0.79$\pm$0.11} & {0.68$\pm$0.11}&{0.68$\pm$0.10} & {13.57$\pm$15.72} & {20.63$\pm$13.90} & {26.11$\pm$9.68} & {19.95$\pm$11.33}\\
		\hline
		{URes-Net\cite{estienne2019u}}&{0.91$\pm$0.04}&{0.72$\pm$0.15} & {0.67$\pm$0.07}&{0.75$\pm$0.08} & {6.80$\pm$2.76} & {11.92$\pm$9.18} & {18.56$\pm$8.53} & {12.18$\pm$12.87}\\
		\hline
		{ProRSeg}&{0.94$\pm$0.02}&{0.86$\pm$0.08} & {0.78$\pm$0.07}&{0.82$\pm$0.05} &{5.69$\pm$1.72}&{7.00$\pm$5.14} & {12.11$\pm$5.30}&{8.11$\pm$3.54}\\
		\hline
		ProRSeg$^{\ddagger}$ &{0.95$\pm$0.009}&{0.91$\pm$0.02} & {0.83$\pm$0.02}&{0.85$\pm$0.04} &{5.52$\pm$0.85}&{6.49$\pm$3.91}&{10.70$\pm$1.20}&{7.35$\pm$2.50}\\
		\hline
		
	\end{tabular}
	}
\end{table}

\subsection{Segmentation consistency with varying organ configurations}
\begin{figure*}
	\begin{center}
        \includegraphics[width=0.8\columnwidth,scale=0.1]{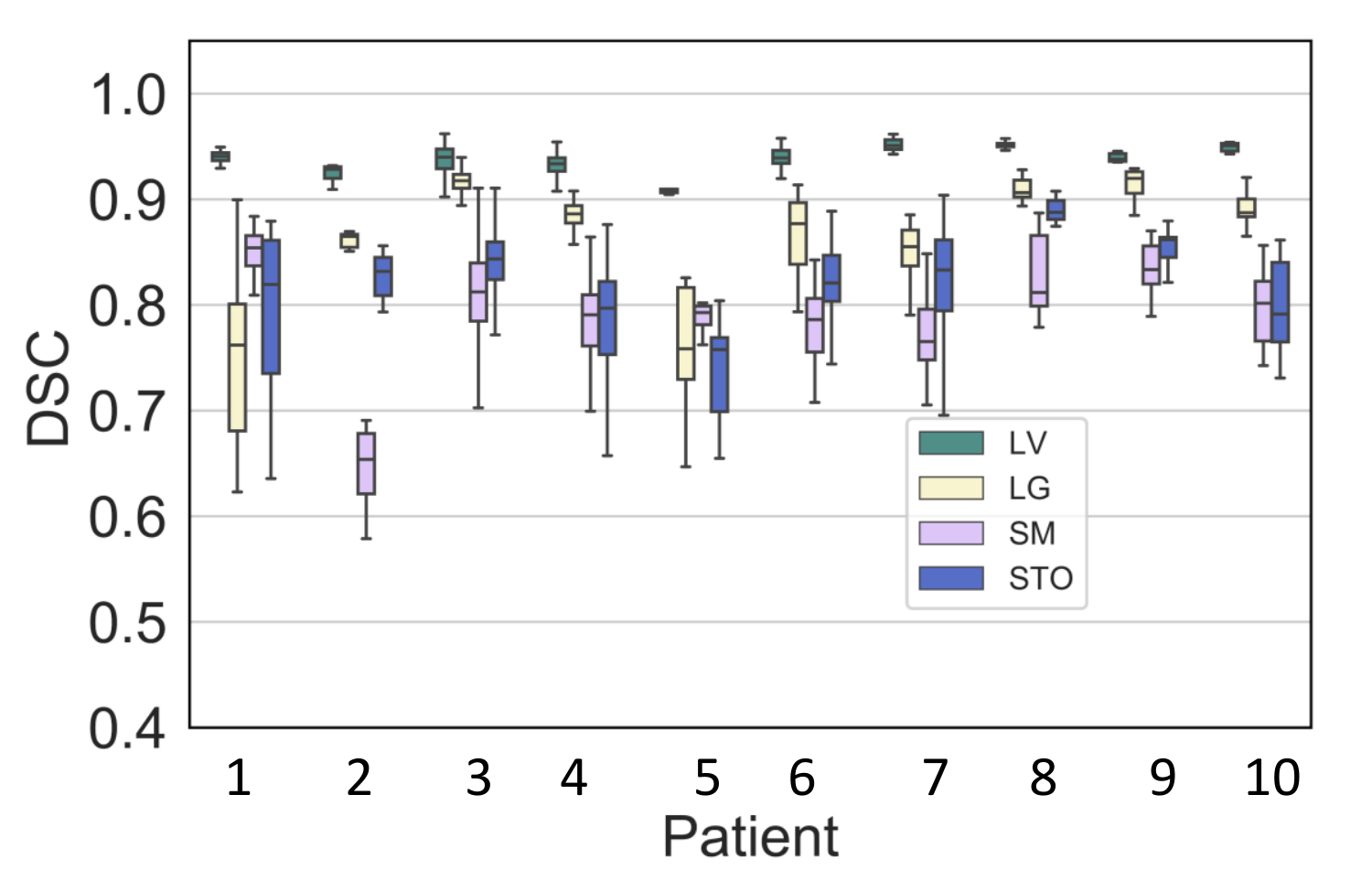}
		\caption{\small Segmentation consistency for all analyzed patients measured using all possible patient-specific pairs (any prior treatment fraction to a current fraction for a given patient) for producing GI OAR segmentations.
			 \label{fig:robust_analysis}}
		\end{center}
\end{figure*}

\begin{figure*}
	\begin{center}
        \includegraphics[width=1.0\columnwidth,scale=0.1]{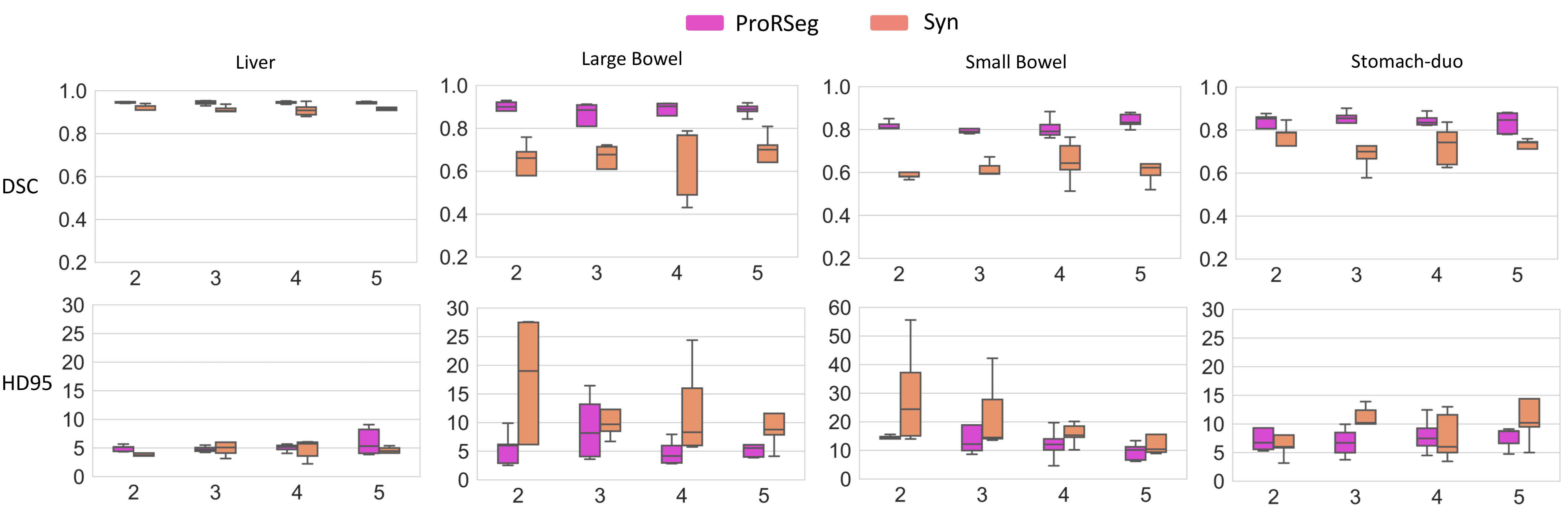}

		\caption{\small Longitudinal variability of segmentation accuracy (DSC and HD95) for ProRSeg and SyN when applied for sequential alignment of treatment fractions as used in the clinic.  
	 \label{fig:const_FX_dsc_hd}}
		\end{center}
\end{figure*}

Figure \ref{fig:robust_analysis} shows the DSC variability for each patient when using all possible combination of treatment fraction MRIs as target and moving image pairs, instead of just aligning consecutive treatment fraction MRIs. This test was performed to evaluate the robustness of segmentation to anatomic configuration of the prior moving image. 
The median CV$_{DSC}$ was under 6\% for all organs with lowest CV$_{DSC}$ observed for the liver (median of 0.45\%, inter-quartile range [IQR] of 0.31\% to 0.63\%) and the highest CV$_{DSC}$ observed for small bowel (median of 4.54\%, IQR of 3.77\% to 5.26\%). Stomach-duodenum (median of 4.04\%, IQR of 3.77\% to 5.26\%) had the second highest CV$_{DSC}$ and large bowel had relatively smaller variation (median of 1.39\%, IQR of 0.61\% to 2.66\%) than both small bowel and stomach-duodenum. The highest overall variation for all organs (combined average of 6.60\% was observed for patient P4 (see Supplemental Table 4) due to large variability in the segmentation of stomach-duodenum (CV$_{DSC}$ of 11.02\%). This specific patient MRI depicted appearance variability due to differences in stomach filling in one of the treatment fractions (\textcolor{black}{3rd row of Figure \ref{fig:seg_overlay_MRI}}). All the remaining patients were treated on empty stomach. Four patients had a CV$_{DSC}$ exceeding 5\% for small bowel and stomach-duodenum, 2 such patients for large bowel, and none for liver. 
\\
Figure \ref{fig:const_FX_dsc_hd} depicts the variability in segmentation accuracies measured using DSC and HD95 across the treatment fractions for the GI OARs. Results produced by iterative deformable image registration using SyN\cite{avants2008symmetric} is also shown for comparison purposes. As shown, ProRSeg shows smaller variability in the segmentation accuracies across the treatment fractions for the analyzed patients compared to the SyN method. Kruskal-Wallis tests of the segmentation accuracies computed across the different fractions showed no difference in DSC (liver: p=0.23, large bowel: p=0.88, small bowel: p = 0.18, stomach: p = 0.46) and HD95 (liver: p = 0.45, large bowel: p = 0.83, small bowel: p = 0.67, stomach: p = 0.65) with ProRSeg method. These results show that ProRSeg generates consistent GI OAR segmentations across the treatment fractions.   

\subsection{Consistency and smoothness of MR-MR DIR}
ProRSeg produced smooth deformations, which were within the accepted range of 1\% of the folding fraction\cite{zhao2019recursive,deVos2019MedIA} (Table~\ref{tab:jacb}). Higher values of $J_{sd}$ and the folding fraction compared to SyN\cite{avants2008symmetric} and Voxelmorph\cite{balakrishnan2019voxelmorph} are expected because ProRSeg allows for more deformation needed to better align the GI organs. The coefficient of variation for the displacement in three directions (x, y, and z) for small bowel, large bowel, and the stomach-duodenum are shown in Table~\ref{tab:jacb}. Liver was excluded in this analysis because only stomach-duodenum, small and large bowel exhibit large deformations and appearance changes. As shown, ProRSeg resulted in the least coefficient of variation in the measured displacements for all three organs, which indicates its ability to produce more consistent registrations. The organ displacements measured in all three directions using ProRSeg and other methods are shown in Supplementary Table 5 and Supplementary Figure 1. ProRSeg measured displacement was the largest for small bowel (x median of 4.92 mm, inter-quartile range [IQR] of 2.31 mm to 8.80 mm; y median of 3.21 mm, IQR of 1.58 mm to 5.48 mm; z median of 4.06 mm, IQR of 2.67 mm to 5.57 mm).

\begin{table} [ht]
\centering{\caption{ Mean and standard deviation of MR-MR DIR smoothness ($J_{sd}$ and $\left|J_\phi \right| \leq 0$ \%) and consistency (CV of displacement). CV per patient is ratio of standard deviation in displacements to the mean displacements for each patient. } 
\label{tab:jacb} 

	\scriptsize
    \begin{tabular}{|c|c|c|c|c|c|c|c|c|c|c|c|} 
		\hline 
        \multirow{2}{*}{Method} & \multirow{2}{*}{J$_{sd}$} & \multirow{2}{*}{$\left| J_{\phi}\right|$} & \multicolumn{3}{c|}{LG Bowel (CV \%)} & \multicolumn{3}{c|}{SM Bowel (CV \%)} & \multicolumn{3}{c|}{Stomach-Duo (CV \%)} \\
        \cline{4-12} 
        {} & {}  & {} & {$CV_{x}$} & {$CV_{y}$} & {$CV_{z}$} & {$CV_{x}$} & {$CV_{y}$} & {$CV_{z}$} & {$CV_{x}$} & {$CV_{y}$} & {$CV_{z}$}  \\
        \hline
		\multirow{2}{*}{SyN\cite{avants2008symmetric}}&{0.04}&\multirow{2}{*}{0.00} & {1.52} & {1.16} & {1.37} & {1.37} & {1.29} & {1.56} & {1.30} & {1.32} & {1.29}\\
		{}&{0.01}&{} & {0.25} & {0.06} & {0.13} & {0.10} & {0.11} & {0.22} & {0.18} & {0.20} & {0.12}\\
		\hline
        \multirow{2}{*}{Voxmorph\cite{balakrishnan2019voxelmorph}}&{0.06}&\multirow{2}{*}{0.00}& {0.94} & {0.89} & {1.00} & {0.84} & {0.85} & {0.92} & {0.78} & {0.78} & {0.92}\\
        {}&{0.01}&{}& {0.05} & {0.14} & {0.42} & {0.17} & {0.06} & {0.42} & {0.10} & {0.18} & {0.51}\\
		\hline
		
		\multirow{2}{*}{URes-Net\cite{estienne2019u}}&{0.09}&\multirow{2}{*}{0.00}& {0.81} & {0.80} & {0.81} & {0.83} & {0.80} & {0.75} & {0.76} & {0.77} & {0.83}\\
        {}&{0.03}&{}& {0.21} & {0.16} & {0.30} & {0.11} & {0.17} & {0.29} & {0.14} & {0.19} & {0.38}\\
		\hline
		
		\multirow{2}{*}{ProRSeg}&{0.18}&\multirow{2}{*}{0.071} & {0.71} & {0.81} & {0.75} & {0.80} & {0.80} & {0.68} & {0.75} & {0.73} & {0.81}\\
		{}&{0.03}&{} & {0.20} & {0.15} & {0.29} & {0.03} & {0.13} & {0.27} & {0.04} & {0.16} & {0.36}\\
		\hline
		
	\end{tabular}
	}
\end{table}

\subsection{Ablation experiments:}
\begin{figure*}
	\begin{center}
        \includegraphics[width=0.9\columnwidth,scale=0.1]{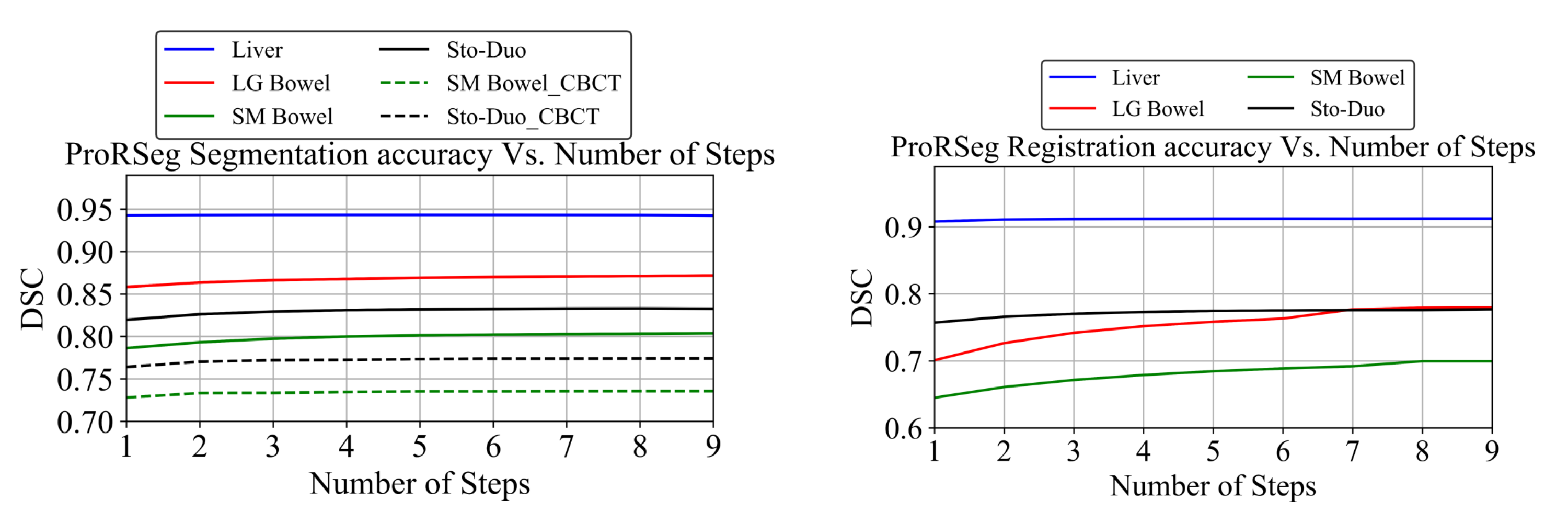}
		\caption{\small Impact of number of CLSTM steps used for RRN and RSN on segmentation accuracy. 
			 \label{fig:acc_vs_step}}
		\end{center}
\end{figure*}

\begin{figure*}
	\begin{center}
        \includegraphics[width=1.0\columnwidth,scale=0.1]{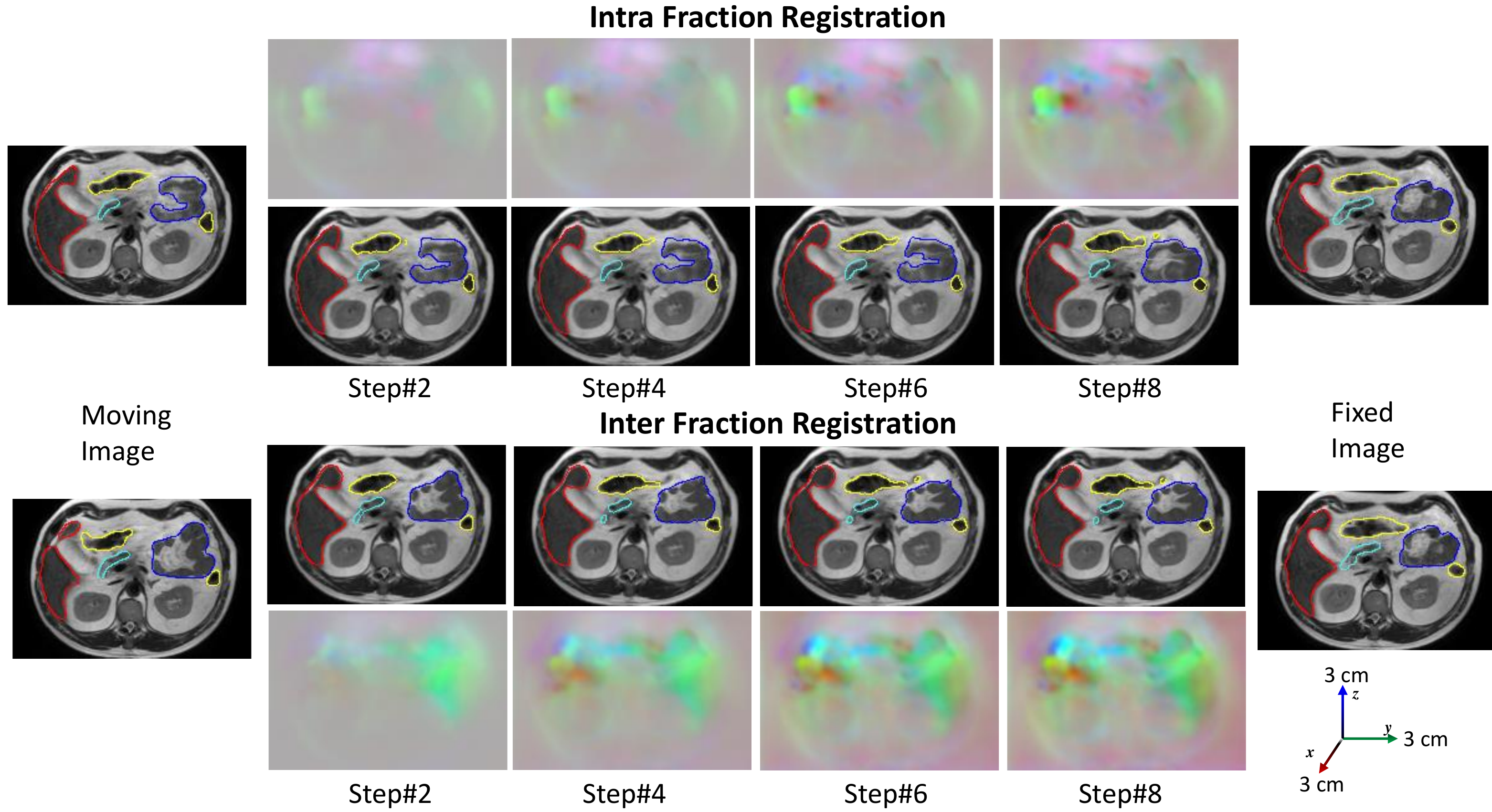}
		\caption{\small Progressive deformation with segmentations produced with ProRSeg shown for aligning intra-fraction (pre-treatment and post-treatment MRI after fraction 1) and inter-fraction (pre-treatment fraction 1 to pre-treatment fraction 2) for a representative patient. 
			 \label{fig:grad_def}}
		\end{center}
\end{figure*}

Accuracies computed by combination of the losses and network design are shown in Table~\ref{tab:ablation}. As shown, segmentation produced by RRN was less accurate than RSN based segmentation, clearly indicating the utility of combining registration with segmentation as a multi-tasked network. Furthermore, removing the segmentation consistency loss reduced the accuracy. Finally, exclusion of spatially aligned prior provided by the RRN to aid the RSN network reduced the accuracy but this prior was less critical than the segmentation loss. Conversely, inclusion of both segmentation consistency and spatially aligned priors resulted in a clear improvement in accuracy for difficult to segment OARs such as the small and large bowel and stomach-duodenum. 
\\
Figure~\ref{fig:acc_vs_step} shows the segmentation accuracy changes due to increasing number of CLSTM steps used in the RRN and RSN networks. As shown, when using RRN to generate segmentations, there was no benefit in increasing the number of CLSTM steps beyond 4. On the other hand, increasing the CLSTM steps in the RSN continued to improve segmentation accuracy for small and large bowel. Figure~\ref{fig:grad_def} shows segmentations produced with increasing number of CLSTM steps for a representative case in both the intra-fraction and inter-fraction registration scenario. As seen, the displacements or DVF are progressively refined with marked displacements occurring during at different steps for the various organs.

\begin{table} [htb]
\centering{\caption{Ablation experiments performed using MRI for GI OAR segmentation.}
\label{tab:ablation} 
	\scriptsize
	\begin{tabular}{|c|c|c|c|c|c|c|} 
		\hline 
	    \multicolumn{3}{|c|}{Method}&\multirow{2}{*}{Liver}&\multirow{2}{*}{LG Bowel}&\multirow{2}{*}{SM Bowel}&\multirow{2}{*}{Stomach} \\ \cline{1-3}
		{Seg consistency} & {Spatial prior} & {Reg-based seg} & & & & \\ 
		\hline
		{$\times$} & {$\checkmark$} & {$\checkmark$} & {0.91$\pm$0.03} & {0.75$\pm$0.13} & {0.69$\pm$0.06}  & {0.76$\pm$0.08} \\
		\hline
		{$\checkmark$}&{$\checkmark$}&{$\checkmark$}& {0.91$\pm$0.03}&{0.78$\pm$0.11} & {0.70$\pm$0.08}&{0.78$\pm$0.06}\\
		\hline
		{$\times$} & {\textcolor{black}{$\checkmark$}} & {$\times$} & {0.93$\pm$0.02}&{0.82$\pm$0.08} & {0.74$\pm$0.08}&{0.78$\pm$0.11} \\
		\hline
		{$\checkmark$ } & {$\times$}& {$\times$}& {0.93$\pm$0.02}&{0.84$\pm$0.08} & {0.76$\pm$0.07}&{0.79$\pm$0.11} \\
		\hline
		{$\times$} &{\textcolor{black}{$\times$}} &{ $\times$}&{0.91$\pm$0.04}&{0.74$\pm$0.11} & {0.67$\pm$0.08}&{0.74$\pm$0.16} \\
		\hline
		{$\checkmark$}&{$\checkmark$}&{$\times$}&{0.94$\pm$0.02}&{0.86$\pm$0.08} & {0.78$\pm$0.07}&{0.82$\pm$0.05}\\
		\hline
		
	\end{tabular}
	}
\end{table}

\subsection{Application of ProRSeg for CBCT OAR segmentation}
\begin{figure*}[tb]
	\begin{center}
        \includegraphics[width=0.5\columnwidth,scale=0.1]{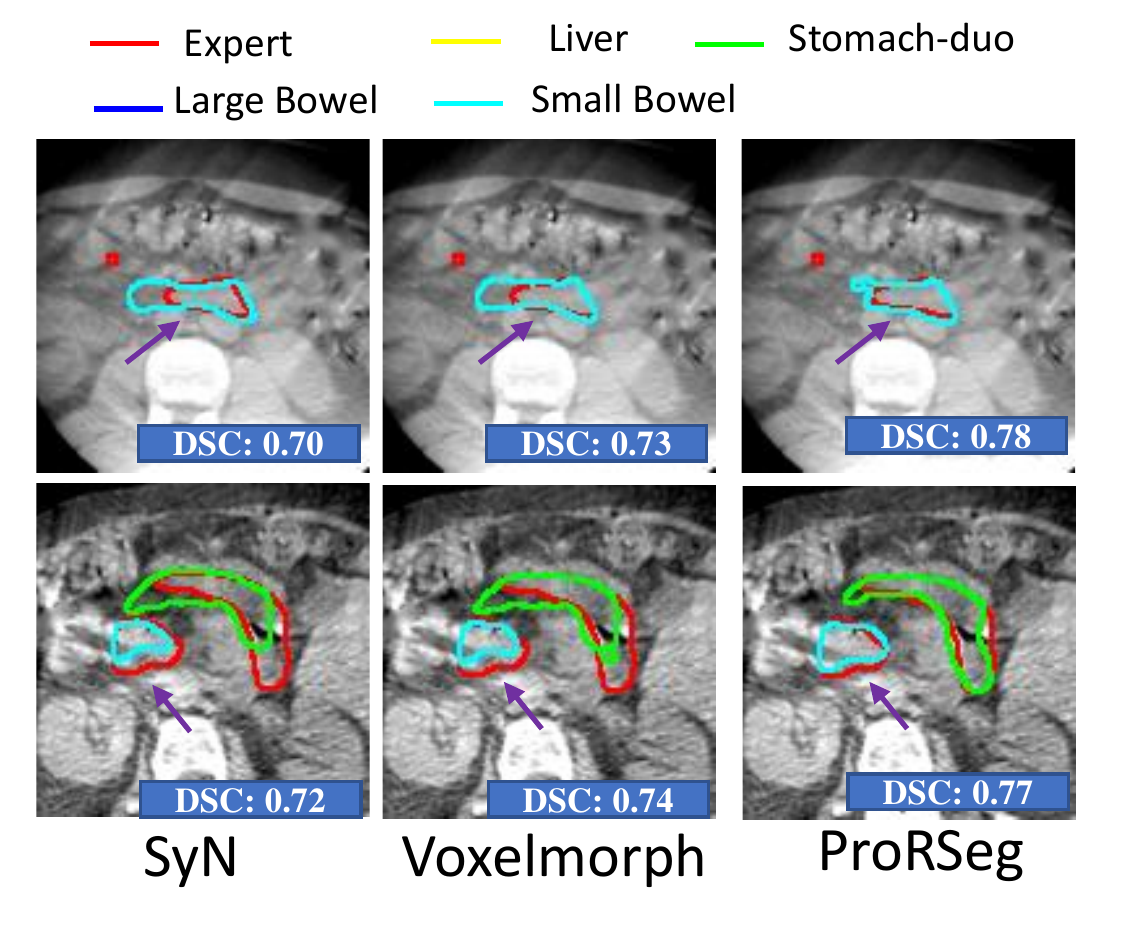}
		\caption{\small Segmentations from CBCT produced on two representative patients using the various methods.
			 \label{fig:seg_overlay_CBCT}}
		\end{center}
\end{figure*}

We next evaluated whether ProRSeg was able to generate segmentation of stomach-duodenum and small bowel from CBCT images. Table~\ref{tab:CBCT_seg} shows a comparison of segmentation accuracies against the SyN~\cite{avants2008symmetric}, Voxelmorph~\cite{balakrishnan2019voxelmorph}, and a previously published method using this same dataset~\cite{han2021deep} and which combined deep learning to learn the momentum parameters to drive the LDDMM. ProRSeg was more accurate than all other methods including Han et al~\cite{han2021deep}. Representative examples from two patients showing segmentations produced by the various methods are in Figure~\ref{fig:seg_overlay_CBCT}. As shown, ProRSeg closely followed expert delineation compared to SyN~\cite{avants2008symmetric} and Voxelmorph~\cite{balakrishnan2019voxelmorph}. 

\begin{table} [h]
\centering{\caption{Segmentation accuracy of various methods applied to CBCT scans. SM Bowel: Small bowel; Stomach-Duo: Stomach duodenum. $^{*}$ HD95 results were not reported.}
\label{tab:CBCT_seg} 
	\scriptsize
	\begin{tabular}{|c|c|c|c|c|} 
		\hline 
		\multirow{2}{*}{Method} & \multicolumn{2}{c|}{DSC $\uparrow$} & 
        \multicolumn{2}{c|}{HD95 mm $\downarrow$} \\
        \cline{2-5} 
		{}&{SM Bowel}&{Stomach-Duo}&{SM Bowel}&{Stomach-Duo} \\
		\hline
		{SyN\cite{avants2008symmetric}}&{0.55$\pm$0.04}&{0.67$\pm$0.03} & {15.87$\pm$2.63}&{19.52$\pm$8.82} \\
		\hline
		{Voxelmorph\cite{balakrishnan2019voxelmorph}}&{0.65$\pm$0.04}&{0.73$\pm$0.03} & {11.43$\pm$3.03}&{13.35$\pm$1.81} \\
		\hline
		{Han et.al\cite{han2021deep}$^{*}$}&{0.71$\pm$0.11}&{0.76$\pm$0.11} & {NA}&{NA} \\
		\hline
		{ProRSeg}&{0.74$\pm$0.02}&{0.77$\pm$0.03} & {10.05$\pm$2.67}&{9.68$\pm$2.67} \\
		\hline
		
	\end{tabular}
	}
\end{table}

\subsection{Proof of principle application of ProRSeg to compute accumulated dose to GI OARs}
\begin{figure*}[htb]
	\begin{center}
        \includegraphics[width=0.85\columnwidth,scale=0.1]{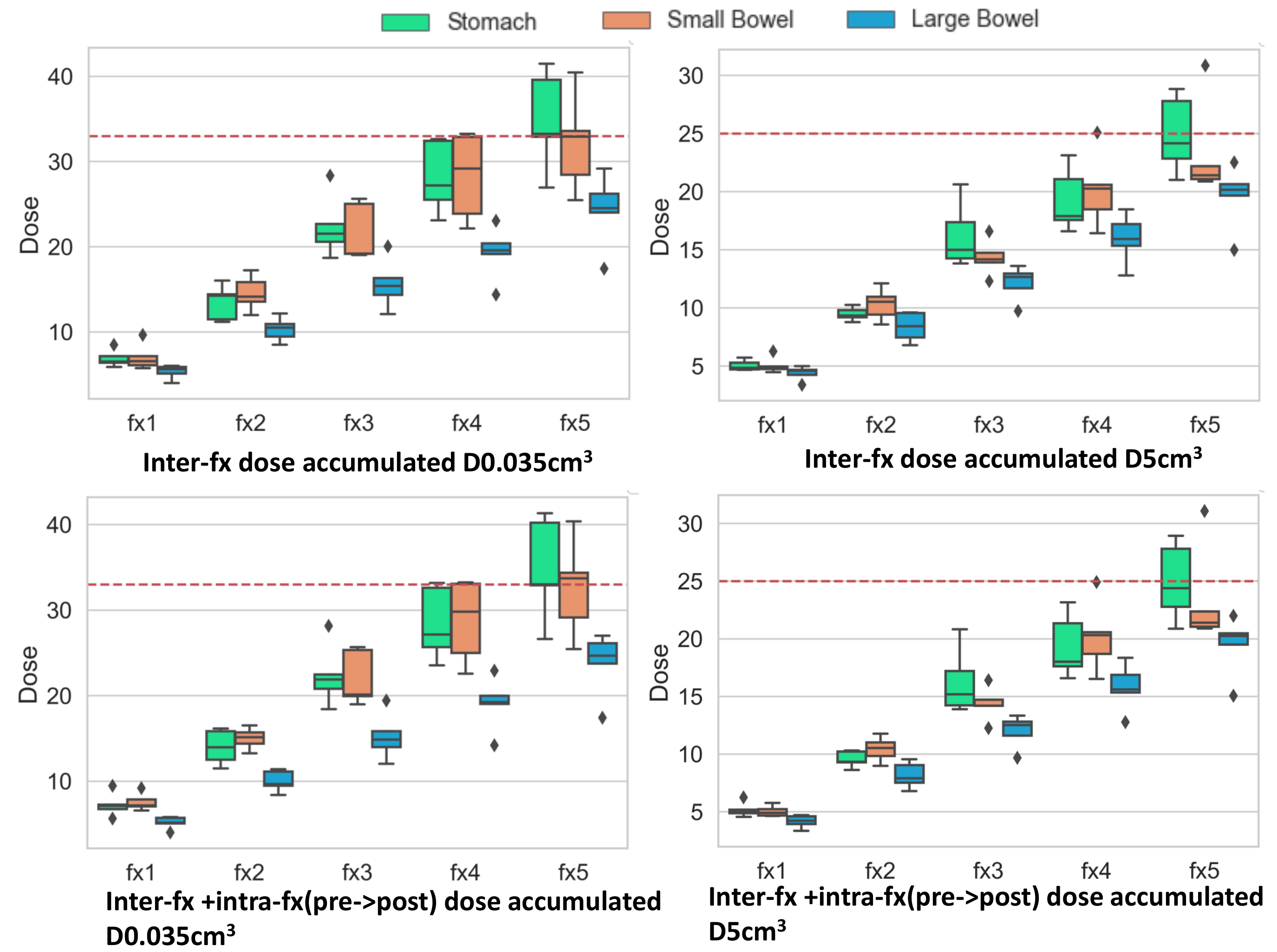}
		\caption{\small Box plots showing accumulated dose metrics D0.035cm$^3$(left) and D5cm$^3$ (right) for stomach duodenum, small bowel and large bowel from interfraction (top row) and inter +  intrafraction (bottom row) accumulated dose of all patients for all 5 fractions. Dotted horizontal line represents the instituitional constraint of 33 Gy (left) for D0.035cm$^3$ and 25 Gy (right) for D5cm$^3$.
			 \label{fig:accumu_dose}}
		\end{center}
\end{figure*}

Figure \ref{fig:accumu_dose} shows the accumulated dose over the course of 5 fractions to the GI OARs without and with intra-fraction dose accumulation. Dose accumulation was performed by sequential alignment of the treatment fraction images and daily fraction doses (scaled to 5 fractions). DVF after each deformation was used to interfractionally accumulate doses for 5 patients who had daily dose maps available from online replanning. 
Intra-fraction dose accumulation was accomplished by aligning the pre-treatment MRI with the post-treatment MRI taken after completion of treatment on the same day. The adaptive plan generated on the pre-treatment MRI in each fraction was copied to the post-treatment MRI and the doses were recalculated, which was then used to compute the intrafraction accumulated doses. 
\\
The institutional dose constraint $D_{max}$ or $D_{0.035cm^{3}} \leq 33 Gy$ and $D_{5cm^{3}} \leq 25 Gy$ are also shown (dotted red lines) in Figure \ref{fig:accumu_dose}. Accumulated dose showed dose violation for stomach-duodenum in four out of 5 patients (Supplemental Table 6). Two patients exceeded both dose constraints for (P2 D$_{0.035cm^{3}}$ = 41.3 Gy, D$_{5cm^{3}}$ = 28.9 Gy; P4 D$_{0.035cm^{3}}$ = 40.2 Gy, D$_{5cm^{3}}$ = 27.8 Gy). Three out of the five patients also violated D$_{0.035cm^{3}}$ dose constraints for the small bowel at fraction 5. However, despite the violation of dose constraints, treatment was well-tolerated in these patients with only one patient (P1) experiencing Grade 1 (mild abdominal pain) acute and late abdominal toxicity\cite{alamRO2021}. Comparison of the accumulated doses for the same patients with LDDMM method used in our prior study\cite{alamRO2021} showed that our method produced a higher estimate of the accumulated doses in general. However, it is difficult to verify the dosimetric accuracy of the individual methods at a voxel-level due to lack of known landmarks to measure target registration error. Also, the prior study\cite{alamRO2021} uses manual segmentations of the OARs in both moving and target images for alignment, which makes comparison of the two methods using volume overlap measures such as DSC and HD95 meaningless.

\section{Discussion}
In this study, we developed and evaluated a multi-task deep registration and segmentation network called ProRSeg to simultaneously segment and deformably align MRI scans longitudinally during radiation treatment course. ProRSeg performs progressive alignment of images as well as refines segmentations progressively by computing dense pixel-level inference using 3D CLSTM implemented into the encoders of registration and segmentation networks. Our approach shows clear accuracy gains compared to segmentation only 3DUnet\cite{ronneberger2015u}, registration-based segmentation using iterative\cite{avants2008symmetric} and deep learning\cite{balakrishnan2019voxelmorph}, and a current simultaneous registration-segmentation method\cite{estienne2019u}. ProRSeg was also applicable to segmenting the more challenging CBCT scans and showed a slightly better accuracy than a current deep learning based LDDMM method\cite{han2021deep} using the same dataset. Ability of ProRSeg to segment on both MRI and CBCT broadens its applicability for radiation treatment planning. 
\\
ProRSeg also produced consistent segmentations and registrations as shown by low CV$_{DSC}$ and low coefficient of variation in the computed displacements for organs compared to other methods. Patient-specific analysis of segmentation accuracy variations due to differences in the anatomical configuration of the prior (or moving images) showed that ProRSeg produced variations within a maximum of 10\%. The larger variations were observed for more complex organs such as the small bowel and stomach-duodenum. At least one patient exhibited differences in stomach volume and appearance due to stomach contents between treatment fractions. In this regard, appearance and anatomic variabilities not encountered in training were difficult to handle in the testing. 
\\
Our analysis showed that ProRSeg produced reasonably accurate GI OAR segmentations exceeding a DSC of 0.80 even for challenging organs such as the small bowel. ProRSeg accuracies are better or comparable to prior published studies applied to different datasets used for MR-Linac treatments\cite{zhang2022,zhang2020,Fu2018}. It is notable that these published methods required ensembling of segmentations\cite{zhang2020} from prior treatment fractions or user editing to drive semi-automated semgentations\cite{zhang2022}. ProRSeg does not require user editing or ensembling. Nevertheless, given differences in datasets, number of training and testing sets, and the way in which the organs were segmented, it's hard to make a head to head comparison of these methods. For example, study by Fu et.al\cite{Fu2018} separated stomach and duodenum into two distinct structures but combined small and large bowel into one structure, whereas we combined stomach and duodenum as one structure but separated the small and large bowels into two distinct structures, consistent with the treatment planning requirements at our institution and  others\cite{zhang2022, ding2022}. Consistent with the findings in a prior study by Zhang et.al\cite{zhang2020}, incorporating prior knowledge from multiple preceding fractions as an ensemble segmentation improved the accuracy for all organs including the small bowel in the later treatment fractions. We will provide our model in GitHub to enable side-by-side comparison by other works upon acceptance for publication. 
\\
Our results are also consistent with prior multi-task methods\cite{xu2019deepatlas,jiang2022CBCTTMI,he2020deep,ZhouMIA2021}, and clearly showed that the inclusion of an additional segmentation network resulted in a higher accuracy than the registration-based segmentation alone. Furthermore, ablation tests clearly showed that spatially aligned priors provided by the registration increased accuracy of segmentation. Finally, incorporating supervised segmentation losses, which is easier to obtain than DVF as ground truths also improved accuracy.   
\\
In addition to segmentation, our method also showed the ability to deformably align images and preliminary feasibility to compute dose accumulation to organs. The computed displacements for organs showed largest median displacement exceeding 4mm for small bowel, which is far below the computed displacements of 10 mm reported using the LDDMM method when applied to a subset of the same patients in a prior study\cite{alamRO2021}. We believe the differences resulted from the additional number of patients included in our study as well as from the averaging of the intra and inter-fraction displacements when computing the overall median displacements. Importantly, our analysis of the dose accumulation showed that ProRSeg measured dose violations were consistent with the findings using LDDMM, albeit ProRSeg produced a higher estimate of accumulated dose for the same patients than LDDMM\cite{alamRO2021} (see Supplementary Table 6). However, it is difficult to assess the voxel-level registration accuracy of either method in order to ascertain the dose accumulation accuracy due to lack of known and visible landmarks to measure target registration error, which also represents a limitation of our study. Robust estimate of TRE for these organs would potentially require synthesizing digital phantom with known landmarks, which was not in the scope of the current study. Also, the prior study\cite{alamRO2021} used manual segmentations of OARs in both moving and fixed image volumes for computing the alignment, which makes comparison of these two methods using volume overlap measures meaningless.  
\\
Our study is limited by lack of set aside testing and lack of large training set for further improving and evaluating the accuracy of the method as well as by the lack of well defined landmarks for measuring target registration error to evaluate registration. Nevertheless, our analysis indicates ability to perform both consistently accurate segmentation and dose accumulation on pancreatic cancer patients using a computationally fast method, thus allowing the use of accumulated doses for potential treatment adaptation in place of the currently used conservative dose constraints.

\section{Conclusion}
A multi-tasked, progressive registration segmentation deep learning approach was developed for segmenting upper GI organs from MRI. Our approach showed ability to produce consistently accurate segmentations and consistent deformable image registration of longitudinal treatment MRI. It was also applicable for segmenting GI organs from cone-beam CT images.

\section*{References}
\bibliographystyle{model2-names.bst}
\bibliography{bibliography}

\end{document}